\documentclass[preprint,aps,draft]{revtex4}
\usepackage{graphicx}
\begin{document}

\title{On the probability distribution function of small scale interplanetary magnetic field fluctuations}
\author{R.~Bruno, B.~Bavassano}
\affiliation{Istituto Fisica Spazio Interplanetario del CNR, 00133 - Roma, Italy}
\author{V.~Carbone, L. Primavera, F. Malara, L. Sorriso-Valvo, P.~Veltri}
\affiliation{Dipartimento di Fisica Universit\`a della Calabria, 87036 Rende (Cs), Italy}

\date{Manuscript version from 5 May 2004}

\begin{abstract}
In spite of a large number of papers dedicated to study MHD
turbulence in the solar wind there are still some simple
questions which have never been sufficiently addressed like:

a)do we really know how the magnetic field vector orientation
fluctuates in space? b) what is the statistics followed by the
orientation of the vector itself? c) does the statistics change as
the wind expands into the interplanetary space?

A better understanding of these points can help us to better
characterize the nature of interplanetary fluctuations and can
provide useful hints to investigators who try to numerically
simulate MHD turbulence.

This work follows a recent paper presented by some of the authors
which shows that these fluctuations might resemble a sort of
random walk governed by a Truncated L\'evy Flight statistics.
However, the limited statistics used in that paper did not allow
final conclusions but only speculative hypotheses. In this work
we aim to address the same problem using a more robust statistics
which on one hand forces us not to consider velocity fluctuations
but, on the other hand allows us to establish the nature of the
governing statistics of magnetic fluctuations with more
confidence.
In addition, we show how features similar to those found in the
present statistical analysis for the fast speed streams of solar
wind, are qualitatively recovered in numerical simulations of the
parametric instability. This might offer an alternative viewpoint
for interpreting the questions raised above.
\end{abstract}
\maketitle

\section{introduction}\label{intro}

The first turbulent model proposed by \cite{kol41} didn't take
into account that the rate of energy transfer along the turbulent
cascade might not be scale--independent (Landau's objection). This
situation, in the framework of a classical Richardson's cascade,
can be described by the fact that smaller and smaller eddies are
less and less space--filling. In other words, turbulence would be
unevenly or intermittently distributed in space. As a matter of
fact, we address this phenomenon as Intermittency. Evidence of
the presence of this phenomenon, when performing a statistical
study on a generic fluctuating field $V(x)$, is that the
probability distribution functions (PDF hereafter) of the
differences $\delta v_l(x)=|V(x+l)-V(x)|$ normalized to the
$\sigma$ of the distribution at scale $l$ do not rescale for
different scales \citep{van75}.
In
particular, the tails of these PDFs become more and more
stretched at smaller and smaller scales. This means that the
wings of the distributions become fatter and fatter. Such a
behaviour implies that, at smaller scales, extreme events become
statistically more probable than if they were normally
distributed.

Intermittency has been found also in the solar wind fluctuations
as demonstrated in the first studies performed by \cite{bur91} in
the outer heliosphere. In particular, this author showed an
unexpected similarity between interplanetary observations on
scales of 1AU and observations on scales of meters obtained for
laboratory turbulence by \cite{ans84}. These results suggested
the universality of this phenomenon, which was independent on
scale. On the other hand, \cite{mar93} and \cite{car95} were the
first authors to study Intermittency in the inner heliosphere. In
particular, the former authors showed the different intermittent
character of fast and slow wind while the latter ones showed a
possible first evidence for the Kraichnan scaling (\cite{kra65})
in a magnetofluid like the solar wind. Since then, several papers
(\cite{ruz95, tu96, hor97, sor99, bru99, pag03, bru03a}) among
others followed these first approaches to the problem of
understanding Intermittency in the solar wind. However, novel
techniques based on the properties of wavelets introduced by
\cite{far90} and firstly used within ordinary fluid dynamics by
\cite{ono00} and within the solar wind context by \cite{vel99},
\cite{bru99} and \cite{bru01}, represented a powerful tool to
finally disclose the nature of intermittent events. In this last
paper, the authors showed that the intermittent event they were
able to single out was a so called current--sheet--associated TD,
as defined by \cite{ho95}. This type of structure is associated
with wind velocity gradients, a rapid magnetic field intensity
change and a reversal of the maximum variance magnetic field
component across the discontinuity. This discontinuity was
interpreted as the interface between two adjacent flux tubes,
i.e. two regions characterized by different plasma and magnetic
field conditions although within the same large scale plasma
region. During these studies, it was also noticed that the jumps
performed by the tips of magnetic and velocity vectors were such
to resemble a typical L\'evy process (\cite{bru04}). This process
is similar to a random walk but the statistics governing the
spatial jumps is characterized by extreme behaviour.
Consequently, the presence of long--range correlations makes the
Gaussian statistics, which governs the Brownian motion, no longer
representative of the physical process. In fact, the spatial
distribution of the directions assumed by these vectors during
the selected time interval was not uniform but rather patchy.
This particular behavior was indicating the presence of
particular directions along which the fluctuating field vector
would roughly remain aligned for a longer time. Then, a rapid and
large jump would characterize the transfer from one patch to
another. These large jumps were recognized to make the
fluctuating field more intermittent. Moreover, the highly
Alfv\'enic character of the selected time interval clearly showed
that propagating modes and coherent structures were both
contributing to the observed turbulence and strengthened the
already proposed view of a turbulence made of a mixture of waves
and structures (\cite{mat90,bru91,tu91,mar93,tu93,kle93}, among
others). In reality, it took a long time to reach this view of
interplanetary MHD turbulence. As a matter of fact, the first
spectra of solar wind fluctuations obtained from the observations
of Mariner 2 in 1962, were interpreted by \cite{col68} as
evidence for the presence of turbulent processes, possibly MHD
turbulence as described by \cite{kra65}. The velocity shear
mechanism proposed by \cite{col68} would be sustained by strong
velocity gradients present in the solar wind and would produce
large scale Alfv\'en waves that would transfer their energy to
smaller and smaller scales through a turbulent process. On the
other hand, \cite{bel71}, looking at the correlation between
velocity and magnetic field fluctuations observed by Mariner 5,
concluded that interplanetary fluctuations were exclusively made
of outward propagating Alfv\'en modes, mostly, of solar origin.
Obviously, these two point of view were in contradiction because
the absence of inward propagating modes would preclude the
development of turbulent cascade of energy to form a spectrum
similar to the one observed by \cite{col68}. In reality,
interplanetary fluctuations are not just Alfv\'en waves. Only
several years later, a careful data analysis performed on the
observations provided by the Helios spacecraft contributed to
make the idea of a possible coexistence of propagating waves and
convected structures accepted (see review by \cite{tu95}).
Moreover, only recently, theoretical efforts (\cite{wu00, vaho04,
vas04} have shown the possibility that propagating modes and
coherent structures might share a common origin within the
general view described by the physics of complexity. Propagating
modes would experience resonances which generate coherent
structures which, in turn, will migrate, interact and eventually
generate new modes. Moreover, \cite{pri03} using a 1--D MHD
numerical simulations based on a pseudo--spectral code, were able
to qualitatively reproduce the radial behavior of magnetic field
and velocity Intermittency observed by \cite{bru03a} in the inner
heliosphere. In particular, they numerically simulated the
propagation of a turbulent Alfv\'enic spectrum in a uniform
background magnetic field. As a matter of fact, coherent
structures were created during the spectral evolution due to the
parametric instability resembling a sort of shocklets or current
sheets. Obviously, the model has strong limitations since while
the 1--D code allows dependence on only 1 spatial coordinate, the
vectors can have all 3 cartesian components. As a consequence,
these results remain at a qualitative level.

Using the idea of \cite{mcc66} as a starting point, \cite{bru01}
proposed the spaghetti--like model in which Alfv\'enic
fluctuations propagate within a convected structure made of
tangled flux--tubes, each of them being characterized by its
local magnetic field and plasma. These structures represent the
correlated part of the signal, unlike from the Alfv\'enic
component which does not conserve spatio--temporal coherence.
Moreover, it was found that the statistics associated to these
fluctuations experienced a strong radial evolution. As a matter
of fact, directional jumps, within fast solar wind, evolved from
a more Gaussian--like statistics at 0.3AU towards a sort of
Truncated L\'evy Flight (\cite{man94} statistics at $0.9$ AU. This
phenomenon suggested that the possible cause of the radial
evolution was the radial progressive depletion of the Alfv\'enic
component of the turbulent fluctuations with respect to the
convected structure. However, the limited statistics due to the
low resolution of plasma data did not allow a more refined
analysis. In this paper we used higher resolution data, about one
order of magnitude higher in frequency, to study in more detail
the structure of the PDFs of directional fluctuations but we had
to limit this study to the only magnetic component of the
fluctuations.

\section{Data analysis}\label{data}


Data used in this work are $6$ sec averages of magnetic field
measurements performed by Helios 2 s/c during its primary mission
to the sun almost three decades ago, in 1976. This is the only
data set available covering the heliocentric distance range
between $0.3$, closest approach to the sun, and 1AU. So far, this
data set has been extremely valuable to study the main physical
mechanisms governing the solar wind turbulence. In particular, it
has been very useful to study the radial evolution of
interplanetary turbulence. As a matter of fact, this data set
contains in--situ observations at different heliocentric
distances of magnetic field and plasma belonging to the same
corotating solar source region of high velocity wind. Thus, the
stationary character of this region allowed studies which could
focus on the radial evolution of solar wind turbulence, providing
extremely important results which have been reviewed in an
excellent paper by \cite{tu95}. In the present analysis we will
refer to this particular high velocity stream and to the low
velocity wind ahead of it observed at two different heliocentric
distances, namely $0.3$ and $0.9$ AU.
\begin{figure}[t]
\vspace*{2mm}
  \caption{\label{fig:fig001} Top panel: solar wind speed profile
  versus time as recorded during the first solar mission of Helios
  2. The solid smooth line represents the heliocentric distance of
  the s/c which varied between 0.97 and 0.29AU.
  Bottom panel: magnetic field intensity profile versus time. The
  whole interval was characterized by the presence of only one
  shock around day 90. Vertical hatched regions identify time
  intervals chosen for this analysis. There are two high speed
  intervals located in the trailing edge of two corotating streams
  and two low speed intervals ahead of them.}
\end{figure}
These time intervals are highlighted in Fig.~\ref{fig:fig001} by
vertical hatched stripes which, from the top to the bottom of the
figure, go across velocity and heliocentric distance in the top
panel and magnetic field intensity in the bottom panel. Each of
the four time intervals lasts 2 days and it has approximately
28800 6sec averages, taking into account data gaps. Temporal
extremes, average heliocentric distance, average wind speed and
field intensity are reported in Table \ref{tab:intervals} for each
interval.

\begin{table}[t]
\centering
  \caption[]
{\label{tab:intervals} Time extremes and average values
characterizing the intervals} \vskip4mm
\renewcommand{\arraystretch}{1.}
\begin{tabular}{cccc}
\hline\noalign{\vskip1mm}time interval & distance [AU] &
$<\mathbf{V}> [km/s]$&  $<\mathbf{B}> [nT]$ \\
\hline
46:00--48:00 & 0.90 & 433 &  6.8 \\
49:12--51:12 & 0.88 & 643 &   6.8 \\
99:12--101:12 & 0.34 & 405 &  28.9 \\
105:12--107:12 & 0.29 & 729 &  42.1 \\
\noalign{\vskip1mm}\hline
\end{tabular}
\end{table}

Just for sake of completeness we like to add that the same
corotating stream was also observed at $0.7$ AU starting around
day 74 but we will omit that time interval since it would be
redundant for the analysis we present here. As already said in
section \ref{intro}, the main goal of this work is the study and
characterization of directional fluctuations. To do so, we start
with plotting the position of the tip of the magnetic field
vector within the reference system of the three coordinate axes
during sub--intervals of only 2000 points within the selected time
periods. These sub--intervals can be considered representative of
each particular interval they refer to. A longer sequence of
points would make impossible, in the graphical format we used, to
recognize differences between different intervals.
\begin{figure}[t]
\vspace*{2mm}
  \caption{\label{fig:fig002} The top panel refers to fast
  wind recorded at 0.3AU while the bottom panel refers to
  fast wind at $0.9$ AU. Each point of both plots
  represents the location of the tip of the
  magnetic field vector for each 6 sec average. These locations have
  then been connected by a black straight line to form a trajectory.
  Moreover, the shadow of this trajectory is also shown on the
  three coordinate planes to better understand the spatial 3D
  configuration.
  In these panels we show only intervals of 2000
  points representing larger intervals of 28800 6--sec
  averages. Values of each component have been normalized
  to the average magnetic field intensity measured within
  the 2000 points interval. The top panel shows a more uniform
  coverage with respect to the bottom panel which shows a sort of
  patchy configuration. In other words, there is some kind of
  evolution during the radial expansion which is dramatically reflected
  in this kind of spatial behaviour.}
\end{figure}
In Fig.~\ref{fig:fig002} we show data relative to fast wind
observed at $0.3$ and $0.9$ AU in the top and bottom panels,
respectively. In this graphical representation, the origin of the
coordinated axes is right in the center of the 3D plot. This
means that, in case of a vector of constant magnitude which
changes direction in a random way, after sufficient time, we
would see a dark sphere centered in the middle of the graph with
radius equal to the vector magnitude. In case directional
fluctuations were concentrated around some particular direction
we would expect the surface of the sphere to be unevenly covered.
In the top panel, the tip of the vector wanders in a random way
on the surface of almost half a sphere without showing any
preferential direction. The enhanced regularity of this surface,
if compared to slow wind in the following, depends on the fact
that, during this high velocity stream, field intensity is quite
constant, as generally expected for a high velocity stream. One
of the main features of this stream is that it is highly
Alfv\'enic as already stressed several times in the available
literature (see for example paper by \cite{bru85}). This fact,
obviously, induces large directional fluctuations on the
direction of the ambient magnetic field. As a consequence, the
dark spot in the figure completely covers half sphere. Obviously,
only half of the sphere is covered since magnetic field polarity
remains generally constant within high velocity streams while it
can flip from positive to negative and vice versa more easily
within slow wind. However, when we reach $0.9$ AU the situation
dramatically evolves. Although the tip of the vector still
wanders on the surface of only half a sphere, it does not cover
this surface completely but leaves out wide areas. The
distribution of the dark spots suggests that the presence of
preferred spatial directions, connected by large and quick jumps
which take only a few data points, begins to emerge as the
heliocentric distance increases.

On the contrary, the evolution we just noticed in fast wind is
much less dramatic within slow wind as shown in the two panels of
Fig.~\ref{fig:fig003}. The top panel, which refers to observations
recorded at $0.3$ AU, clearly shows the presence of two main
directions around which the magnetic field vector fluctuations
mainly cluster. Moreover, as in the previous case, large jumps
connecting one spot to the other are clearly visible. When we
move to $0.9$ AU in the lower panel, we do not notice much of a
difference with respect to the situation encountered in the upper
panel. Fluctuations appear to be generally smaller but the spotty
configuration clearly emerges again.
\begin{figure}[t]
\vspace*{2mm}
  \caption{\label{fig:fig003}
  The top panel refers to slow wind recorded at $0.3$ AU
  while the bottom panel refers to
  slow wind at $0.9$ AU. The format is the same used for
  Fig.~\ref{fig:fig002} and,
  also in this case, we show only 2000 points
  out of 28800 points of each selected time interval. Both configurations
  relative to 0.3 and $0.9$ AU greatly differ from what we observed within
  fast wind. In both cases a patchy configuration
  is clearly visible, meaning that the tip of the vector dawdles
  longer around some particular orientations. At first sight,
  the two panels show similar configurations although fluctuations
  at $0.3$ AU appear to be slightly larger. Definitely, we do not
  observe the same radial evolution noticed for fast wind}
\end{figure}
However, this kind of graphical representation is not sufficient
to give an idea of how the tip of the vector really moves in time
unless complemented by the information we provide in the top
panel of Fig.~\ref{fig:fig004}. In this panel we show the vector
displacement $|\delta\underline{B}(t)|$, normalized to
$<|\underline{B}|>$, between each $\underline{B}(t)$ and an
arbitrary fixed direction which we chose to be the direction of
the first vector $\underline{B}(t_0)$ of the time series. Thus,
following this definition, each individual
$|\delta\underline{B}(t)|$ is given by:
\begin{equation}\label{eq:deltab}\nonumber
|\delta\underline{B}(t)|=\sqrt{\sum_{i=x,y,z}(B_i(t)-B_i(t_0))^2}
\end{equation}
This time sequence, the same used for the top panel of
Fig.~\ref{fig:fig003} which refers to slow wind at $0.3$ AU,
clearly shows small amplitude and high frequency fluctuations
superimposed on a sort of larger amplitude low frequency
background structure. This background structure is characterized
by few large and quick directional jumps. The effect of these
jumps is to move the fluctuating vector from one particular
average direction to another, i.e. from one dark spot to another
(Fig.~\ref{fig:fig003}). This type of information, together with
the 3D graphical representation, gives an idea of how the vector
direction really fluctuates in space and time. Moreover, most of
the time the largest directional jumps are associated to the
largest changes of the field intensity (bottom panel). So, these
two panels suggest that during short time intervals the field can
be characterized by a most probable orientation and a most
probable intensity. In other words, these regions appear to be
distinguishable from each other and the transition from one to
another is through a large rotational jump and a change in the
field intensity. Similar findings have already been reported in a
previous paper (\cite{bru01}) although it was focused on a single
case study and on larger scales. In that same study it was found
that this kind of transition, or border, was a tangential
discontinuity not in pressure balance. The features we notice in
the present study might be TDs as well although we cannot prove
it since we don't have plasma data with the same time resolution
of magnetic field data. If this is the case, the structure we
have seen at larger scales replicates at smaller scales in a kind
of self-similar manner.

Results relative to slow wind at $0.9$ AU are shown in
Fig.~\ref{fig:fig005} in the same format as of the previous
Figure. Although, both panels show fluctuations generally smaller
than the corresponding ones observed at $0.3$ AU, corresponding
features in both profiles are still clearly recognizable. Thus,
radial evolution within slow wind doesn't play much of an
influence on this kind of relationship.
\begin{figure}[t]
\vspace*{2mm}
  \caption{\label{fig:fig004} Data refer to slow wind at $0.3$ AU.
  Top panel: vector displacement $|\delta\underline{B}(t)|$, normalized to
  $<|\underline{B}|>$, between each $\underline{B}(t)$ and an
  arbitrary fixed direction versus time. The arbitrary direction
  was chosen as the direction of the first vector of the time
  series. This kind of graph shows a series of time intervals
  during which the vector displacement tends to remain approximately
  close the average level. These time intervals are interleaved by
  large and quick directional jumps. Moreover, largest jumps often coincide with
  remarkable changes in field intensity as shown in the bottom panel.}
\end{figure}
\begin{figure}[t]
\vspace*{2mm}
  \caption{\label{fig:fig005} Vector displacement
  versus time in the same format as of Fig.~\ref{fig:fig004} relative to
  slow wind at $0.9$ AU. Largest vector displacements (top panel) often
  coincide with large compressive events as shown in the bottom panel.
  As shown in the paper, the statistics of these vector displacements
  is remarkably similar to that shown in the previous Fig.~\ref{fig:fig004}}
\end{figure}
In the next Fig.~\ref{fig:fig006} we show vector displacement for
the time interval recorded at $0.3$ AU, within fast wind in the
same format of the previous two Figures. Directional fluctuations
appear to be very chaotic and not as much structured as we found
in slow wind. Thus, it is certainly more difficult to recognize
structures similar to the ones observed in the previous Figures
and correlate them to the profile of the magnetic field intensity
in the bottom panel. As a matter of fact, we expect to find large
amplitude directional fluctuations within fast wind, especially
close to the sun, because we are aware of the relevant presence
of Alfv\'enic modes in this type of wind. As a consequence, we
believe that these fluctuations mask the correspondence we were
able to highlight within slow wind which, \textsl{a priori},
might be similar to that. Consequently, if the Alfv\'enic modes
had a smaller amplitude we would be able to recognize and relate
similar features in both panels.

As a matter of fact, results shown in Fig.~\ref{fig:fig007}
relative to fast wind data recorded at $0.9$ AU, strongly support
our intuition. The top panel of this Figure shows a well
structured profile, not much different from those observed in
slow wind. This time, several structures noticed in the top panel
can easily be related to companion structures in the bottom
panel. The reason lies in the fact that the amplitude of
Alfv\'enic fluctuations is largely reduced at the Earth's orbit
compared to short heliocentric distances (see the wide related
literature in \cite{tu95}).
\begin{figure}[t]
\vspace*{2mm}
  \caption{\label{fig:fig006} Fast wind at $0.3$ AU. Normalized vector
  displacements versus time in the same format as of Fig.~\ref{fig:fig004} and
  Fig.~\ref{fig:fig005} are shown in the top panel while, normalized
  vector intensities are shown in the bottom panel. The top panel shows
  large fluctuations which are difficult to relate to the profile of the
  magnetic field intensity in the bottom panel although, some corresponding
  events can still be recognized. In this sense, fast wind at $0.3$ AU
  differs from the slow wind samples we already discussed.}
\end{figure}
\begin{figure}[t]
\vspace*{2mm}
  \caption{\label{fig:fig007} Normalized vector displacements and normalized
  vector intensity versus time are shown in the same format as of
  Fig.~\ref{fig:fig004} to Fig.~\ref{fig:fig006} for fast wind at $0.9$ AU.
  Vector displacements shown in the top panel appear to be less chaotic than
  those observed at $0.3$ AU. A sort of underlying structure can be recognized
  and related to field intensity fluctuations shown in the bottom panel.
  In other words, this situation tends to resemble the one encountered
  within slow wind.}
\end{figure}
This qualitative study has to be substantiate with some
quantitative evaluation of the relative importance of these two
components contributing to the observed interplanetary turbulence.
To do so, in the following, we will discuss and compare the
probability distribution functions (PDF) of the directional
fluctuations observed within each time interval. In order to look
for a possible scaling between different PDFs, we have normalized
each $|\delta\underline{B}(t)|$ to the standard deviation
$\sigma$of the relative distribution. Moreover, the maximum
amplitude of each PDF was normalized to $1$. In
Fig.~\ref{fig:fig008} we show PDFs for the fast wind samples
recorded at 0.3 and $0.9$ AU in the top and bottom panels,
respectively. We found that both distributions can be reasonably
well fitted by a double lognormal distribution in the form
reported by eq.~\ref{eq:fit01}.
%
\begin{equation}\label{eq:fit01}\nonumber
P(\xi) = \frac{A_{1}}{\sigma_{1}\xi\sqrt{2\pi}}
\exp\left[-\left(\frac{\ln|\xi/\delta_{1}|}{\sqrt{2}\sigma_{1}}\right)^{2}\right]
 + \frac{A_{2}}{\sigma_{2}\xi\sqrt{2\pi}}
\exp\left[-\left(\frac{\ln|\xi/\delta_{2}|}{\sqrt{2}\sigma_{2}}\right)^{2}\right]
\end{equation}
where $\xi,\delta,\sigma>0$.  The variable $\xi$ stands for the
different $|\delta\underline{B}_i|/\sigma$, one for each bin of
the distribution, $A_i$ is a measure of the area under each curve,
$\delta_i$ is called scale parameter and represents the median,
$\sigma_i$ is the shape parameter. Larger values of $\sigma_i$
push the x--location of the peak of the distribution towards
lower values. Obviously, using a larger number of lognormals
would provide a better fit but, the real conspicuous improvement
is obtained only when we use two lognormals instead of just one.

The two lognormal components have a distinguishable role in the
total PDF. One component accounts for the lognormal tail at large
values of $|\delta\underline{B}|/\sigma$ while the second
component takes care of the smallest values of
$|\delta\underline{B}|/\sigma$. Moreover, taking into account the
values of the parameters reported in Tab.~\ref{tab:paramfast} and
relative to the best fit obtained reaching the minimum $\chi^2$
value for the two distributions at 0.3 and $0.9$ AU, one of the
two lognormals experiences a stronger radial evolution. In
particular, the lognormal that represents fluctuations peaked on
smaller $|\delta\underline{B}|/\sigma$ strongly decreases its
contribution with increasing radial distance from the sun. An
estimate of this evolution can be inferred from the ratio of the
areas $A_1$ and $A_2$ below each curve. While at $0.3$ AU the
probability ratio $A_2/A_1\simeq 0.52$, at $0.9$ AU it drops to
$\simeq 0.22$. Consequently, the contribution of the smaller PDF
to the whole PDF varies from $34\%$ at $0.3$ AU to $18\%$ at
$0.9$ AU. All the other parameters do not experience a similar
radial variation and this behavior reflects in a depletion of the
left--hand tail of the total PDF. However, it is worth noticing
that values of both $\delta$'s are located at somewhat larger
values for the sample referring at $0.3$ AU, suggesting that
fluctuations are generally larger when closer to the sun.
\begin{table*}[t]
\centering
  \caption[]
{\label{tab:paramfast}Fast Wind: Values of the Parameters Obtained
from the Fit of the PDF relative to vector displacements}
\vskip4mm
\renewcommand{\arraystretch}{1.2}
\begin{tabular}{cccccccc}
\hline\noalign{\vskip1mm} Distance~AU & $r^2$ & $A_1$ &
$\sigma_1$ & $\delta_1$ & $A_2$
& $\sigma_2$ & $\delta_2$  \\
\hline $0.3$ & $0.998$ & $0.85\pm0.13$  &  $0.71\pm0.03$  &
$0.95\pm0.02$
&  $0.44\pm0.14$  &  $1.08\pm0.04$  &   $0.60\pm0.13$  \\
$0.9$ & $0.998$ & $0.89\pm0.24$  &  $0.74\pm0.03$  & $0.80\pm0.02$
&  $0.20\pm0.25$  &  $0.98\pm0.09$  &   $0.52\pm0.32$  \\
\noalign{\vskip1mm}\hline
\end{tabular}
\end{table*}

\begin{table*}[t]
 \centering
  \caption[]
{\label{tab:anglefast}Fast Wind: Values of the Parameters Obtained
from the Fit of the PDF relative to angular fluctuations}
\vskip4mm
\renewcommand{\arraystretch}{1.2}
\begin{tabular}{cccccccc}
\hline\noalign{\vskip1mm} Distance~AU & $r^2$ & $A_1$ &
$\sigma_1$ & $\alpha_1$ & $A_2$
& $\sigma_2$ & $\alpha_2$  \\
\hline $0.3$ & $0.998$ & $8.47\pm0.76$  &  $0.72\pm0.02$  &
$9.24\pm0.14$
&  $4.77\pm0.84$  &  $1.16\pm0.03$  &   $6.28\pm0.81$  \\
$0.9$ & $0.999$ & $4.62\pm0.78$  &  $0.75\pm0.03$  & $4.67\pm0.09$
&  $1.87\pm0.82$  &  $1.07\pm0.05$  &   $3.00\pm0.82$  \\
\noalign{\vskip1mm}\hline
\end{tabular}
\end{table*}
Similar conclusions apply to the PDFs relative to the angular
fluctuations $\delta\alpha$ experienced by the vector orientation
shown in Fig.~\ref{fig:fig009}. Obviously, this measure provides
information only about directional fluctuations and is not
influenced by compressive effects that may act on the vector
intensity. As such, information contained in Fig.~\ref{fig:fig009}
are less meaningful than those discussed earlier but, we like to
show this kind of Figure, and the analogous one for slow wind in
Fig.~\ref{fig:fig011}, just for sake of completeness. Moreover,
these distributions have not been normalized to their respective
$\sigma$s since we like to show the effective angular range of
these fluctuations. Also for this fit we report the relative
parameters which are shown in Tab.~\ref{tab:anglefast}. For these
fluctuations the ratio $A_2/A_1$ varies from $56\%$ at $0.3$ AU
to $40\%$ at $0.9$ AU. The PDF is clearly peaked at larger angles
($5.75^\circ$ compared to $2.25^\circ$) at $0.3$ AU and its right
tail reaches values close to $100^\circ$.
\begin{figure}[t]
\vspace*{2mm}
  \caption{\label{fig:fig008} PDFs of vector displacements
  $|\delta\underline{B}|$ normalized to $\sigma$ at 0.3 and $0.9$ AU, for fast wind, are
  shown in the top and bottom panels, respectively. The two thin solid curves
  refer to as many lognormals contributing to form the thick solid curve which
  best fits the distribution. Parameters relative to the fit are reported in
  Tab.~\ref{tab:paramfast}.}
\end{figure}
\begin{figure}[t]
\vspace*{2mm}
  \caption{\label{fig:fig009} PDFs of angular displacements
  $\delta\alpha$ at $0.3$ and $0.9$ AU, for fast wind, are
  shown in the top and bottom panels, respectively. The two thin solid curves
  refer to as many lognormals contributing to form the thick solid curve which
  best fits the distribution. Parameters relative to the fit are reported in
  Tab.~\ref{tab:anglefast}.}
\end{figure}
In Fig.~\ref{fig:fig010} we show the vector displacement
$|\delta\underline{B}|$ normalized to $\sigma$ for slow wind at
$0.3$ and $0.9$ AU in the same format as of Fig.~\ref{fig:fig008}.
The situation within slow wind shows that radial evolution is
almost absent. This can be inferred from the values of the best
fits parameters reported in Tab.~\ref{tab:paramslow} which show
that the contribution of the smaller lognormal can be neglected at
both distances. As a matter of fact, the relative contribution of
the smaller lognormal is between $0.26\%$ at $0.3$ AU and $0.13\%$
at $0.9$ AU, respectively. Moreover, the values of the parameters
relative to the larger lognormal only slightly change between 0.3
and 0.9 AU suggesting that these fluctuations do not evolve much
with radial distance, as expected for slow wind. In summary,
remarkable is the constancy of all parameters inferred from the
fit of the main lognormal, which highlights the absence of radial
evolution. In this case, values of $\delta$'s are considerably
smaller than those obtained for fast wind, confirming that these
fluctuations are generally smaller.
\begin{table*}[t]
\centering
  \caption[] {\label{tab:paramslow}Slow Wind: Values of
the Parameters Obtained from the Fit of the PDF relative to
vector displacements} \vskip4mm
\renewcommand{\arraystretch}{1.2}
\begin{tabular}{cccccccc}
\hline\noalign{\vskip1mm} Distance~AU & $r^2$ & $A_1$ &
$\sigma_1$ & $\delta_1$
& $A_2$ & $\sigma_2$ & $\delta_2$  \\
\hline $0.3$ & $0.998$ & $0.776\pm0.003$  &  $0.959\pm0.009$  &
$0.523\pm0.004$
&  $0.002\pm0.001$  &  $0.437\pm0.095$  &   $0.039\pm0.010$  \\
$0.9$ & $0.998$ & $0.727\pm0.004$  &  $0.934\pm0.005$  &
$0.503\pm0.004$
&  $0.001\pm0.001$  &  $0.532\pm0.722$  &   $0.049\pm0.009$  \\
\noalign{\vskip1mm}\hline
\end{tabular}
\end{table*}

\begin{table*}[t]
\centering
  \caption[]
{\label{tab:angleslow}Slow Wind: Values of the Parameters Obtained
from the Fit of the PDF relative to angular fluctuations}
\vskip4mm
\renewcommand{\arraystretch}{1.2}
\begin{tabular}{cccccccc}
\hline\noalign{\vskip1mm} Distance~AU & $r^2$ & $A_1$ &
$\sigma_1$ & $\alpha_1$
& $A_2$ & $\sigma_2$ & $\alpha_2$  \\
\hline $0.3$ & $0.999$ & $4.14\pm0.68$  &  $1.08\pm0.02$  &
$2.74\pm0.33$
&  $0.31\pm0.66$  &  $0.76\pm0.32$  &   $5.12\pm0.76$  \\
$0.9$ & $0.999$ & $2.68\pm0.02$  &  $0.98\pm0.01$  & $1.89\pm0.01$
&  $0.12\pm0.02$  &  $0.42\pm0.03$  &   $1.27\pm0.02$  \\
\noalign{\vskip1mm}\hline
\end{tabular}
\end{table*}
As already reported for fast wind, we like to show the PDFs
relative to the angular fluctuations as shown in
Fig.~\ref{fig:fig011} and  parameters relative to the best fit
which are shown in Tab.~\ref{tab:angleslow}. Also in this case,
the contribution of the smaller lognormal is much smaller than
within fast wind. As a matter of fact, the ratio $A_2/A_1$ varies
from $7.5\%$ at $0.3$ AU to $4.4\%$ at $0.9$ AU. Moreover, these
PDFs are roughly peaked at the same angle ($0.75^\circ$) although
the right tail at $0.3$ AU reaches larger values and, do not show
any noteworthy radial dependence.
\begin{figure}[t]
\vspace*{2mm}
  \caption{\label{fig:fig010} PDFs of vector displacements
  $|\delta\underline{B}|$ normalized to $\sigma$ at 0.3 and $0.9$ AU, for slow wind, are
  shown in the top and bottom panels, respectively. The two thin solid curves
  refer to as many lognormals contributing to form the thick solid curve which
  best fits the distribution. Parameters relative to the fit are reported in
  Tab.~\ref{tab:paramslow}. The smaller lognormal is almost superfluous since
  its contribution to the final fit is negligible.}
\end{figure}
\begin{figure}[t]
\vspace*{2mm}
  \caption{\label{fig:fig011} PDFs of angular displacements
  $\delta\alpha$ at 0.3 and $0.9$ AU, for slow wind, are
  shown in the top and bottom panels, respectively. The two thin solid curves
  refer to as many lognormals contributing to form the thick solid curve which
  best fits the distribution. Parameters relative to the fit are reported in
  Tab.~\ref{tab:angleslow}. Also in this case, the contribution of the smaller
  lognormal to the final fit is negligible.}
\end{figure}
%

\subsection{Building artificial interplanetary time series}

At this point, we tried to reproduce, from a statistical point of
view, our interplanetary data samples employing a random walk
process governed by a double log--normal statistics acting on the
direction of a unit vector. In other words, the interval of
variability of $|\delta\underline{B}_i|$, as inferred from real
data, was divided in a sufficient number of bins. For each of
them we generated a certain number of values, all equal to the
value represented by the mid point of the bin. The amount of
values generated depended on the corresponding probability
indicated by the double log--normal, which was shaped using the
same parameters we had previously obtained from our best fits and
reported in Tables~\ref{tab:paramfast} and \ref{tab:paramslow}.
These $|\delta\underline{B}_i|$ were then randomly extracted and
used to make the tip of a unit vector, with one end fixed at the
center of a sphere of unit radius, wander on the surface of the
sphere. The direction of the path followed by the tip of the
vector at each step was randomly extracted between $0^\circ$ and
$360^\circ$. In particular, to avoid the effect of the two
singular points at the poles of the sphere, the Cartesian
coordinates of our reference system were rotated after each
extraction in order to have the $x$ axis always coinciding with
the newly extracted direction.
\begin{figure}[t]
\vspace*{2mm}
  \caption{\label{fig:fig012} 3D representation of vector displacements
  relative to artificial data generated by a random--walk whose jumps are
  obey to a double--lognormal whose parameters have been obtained by the best fit
  of real fluctuations. The top and bottom panels refer to fast wind at 0.9 and
  $0.3$ AU, respectively and, have to be compared to analogous plots for real data shown in
  Fig.~\ref{fig:fig002}.}
\end{figure}
\begin{figure}[t]
\vspace*{2mm}
  \caption{\label{fig:fig013} 3D representation of vector displacements
  relative to artificial data generated by a random--walk whose jumps are
  obey to a double--lognormal whose parameters have been obtained by the best fit
  of real fluctuations. The top and bottom panels refer to slow wind at 0.9 and
  $0.3$ AU, respectively and, have to be compared to analogous
  plots for real data  shown in Fig.~\ref{fig:fig003}.}
\end{figure}
Four artificial temporal series of 28800 data points each,
representing interplanetary observations performed at 0.3 and 0.9
AU, for fast and slow wind were built in such a way. Samples of
2000 data points each are plotted in Fig.~\ref{fig:fig012} and
Fig.~\ref{fig:fig013}, for fast and slow wind, respectively. The
only arbitrary imposition we applied on the fast wind sample was
that to keep always the same vector polarity to resemble, as much
as possible, the real situation within fast wind. As a
consequence, we forced these fluctuations to remain within a
solid angle of $2\pi$ aperture. These plots reproduce at some
level the main features that can be observed in
Fig.~\ref{fig:fig002} and Fig.~\ref{fig:fig003}. Fluctuations
appear more intermittent in slow wind but show the largest
evolution, between 0.3 and $0.9$ AU, within fast wind. Obviously,
all the artificial time series we built have, by definition, the
same statistics of real interplanetary data and we omit to show
the PDFs of the relative $|\delta\underline{B}_i|/\sigma$ or
$\delta\alpha$.

On the other hand, we like to show temporal sequences of these
$|\delta\underline{B}(t)|$ relative to a fixed, arbitrary
direction as we did for real data shown in the top panels of
Fig.~\ref{fig:fig004} to ~\ref{fig:fig007}. Only the top panels
have to be considered since, artificial data have been built
keeping the vector intensity constant. Results are quite
satisfactory since we are able to reproduce the typical behavior
observed within both fast and slow wind. In particular, the
transition from the chaotic behavior on the left panel of
Fig.~\ref{fig:fig014}, representing fluctuations at $0.3$ AU,
towards more structured fluctuations on the right panel of the
same Figure, representing fluctuations at $0.9$ AU, is well
reproduced by the artificial time series.
\begin{figure}[t]
\vspace*{2mm}
  \caption{\label{fig:fig014} Vector displacement versus time as measured
  from artificial data referring to fast wind at $0.3$ AU on the left panel and
  at $0.9$ AU on the right panel. These plots should be compared to the top panels
  of Fig.~\ref{fig:fig006} and Fig.~\ref{fig:fig007}, respectively.}
\end{figure}
\begin{figure}[t]
\vspace*{2mm}
  \caption{\label{fig:fig015} Vector displacement versus time as measured
  from artificial data referring to slow wind at $0.3$ AU on the left panel and
  at $0.9$ AU on the right panel. These plots should be compared to the top panels
  of Fig.~\ref{fig:fig004} and Fig.~\ref{fig:fig005}, respectively..}
\end{figure}
Moreover, artificial data reproduce equally well fluctuations
encountered at both heliocentric distances in slow wind. They
show similar structured fluctuations and not relevant differences
between 0.3 and $0.9$ AU.

The last comparison between real and artificial time series that
we like to show refers to the power spectra associated to
fluctuations experienced by the vector components. In order to do
so, we computed the trace of the power spectrum for real and
artificial fluctuations relative to both heliocentric distances
within fast wind. Results are shown in Fig.~\ref{fig:fig016}
where, power spectra relative to real fluctuations are reported
in the left--hand--side panel while, corresponding spectra of
artificial time series are in the right--hand--side panel.
\begin{figure}[t]
\vspace*{2mm}
 \caption{\label{fig:fig016} The left panel shows power spectra obtained
  from the trace of the spectral matrix relative to magnetic field fluctuations
  recorded at 0.3 and $0.9$ AU by Helios 2 in 1976. Each time interval had 2048
  averages of 6--sec each and are the same sub--intervals used for
  Fig.~\ref{fig:fig002}, \ref{fig:fig006}, \ref{fig:fig007}.
  The right panel shows power spectra obtained
  from the trace of the spectral matrix relative to the artificial field fluctuations
  relative to 0.3 and $0.9$ AU, built from a random walk whose jumps obey to a
  double--lognormal distribution. Data relative to $0.3$ AU have been multiplied by
  a factor of $10^2$ to facilitate visual comparison with real data shown in the
  left panel. It is worth to notice that artificial data, besides a general agreement
  with real data, are able to reproduce the bending of the power spectrum observed
  at $0.3$ AU. Straight solid lines indicate the $k^{-5/3}$ Kolmogorov slope }
\end{figure}
The power spectra of the components have been computed via a Fast
Fourier Transform from time series of 2048 data points. The power
spectral densities of the three components have been successively
added up to obtain the trace of the variance matrix which has
been smoothed by averaging adjacent data points within a sliding
window of 5 points. In the same panels we also show as a
reference the slope of the classical Kolmogorov's spectrum.
Spectra shown in the left--hand--side panel are typical spectra
encountered within high velocity streams, as several times
reported in literature (see review by \cite{tu95}). On the other
hand, artificial spectra have been graphically separated by
multiplying the spectrum identified by the label $0.3$ AU by a
factor of $10^2$ to avoid overlapping. As a matter of fact, our
artificial fluctuations have statistically similar amplitudes, no
matter whether we refer to 0.3 or $0.9$ AU since our fluctuations
are confined onto the surface of a sphere of unitary radius.
Unexpectedly, the resemblance is so good that the artificial
spectrum at $0.3$ AU shows a bending similar to the one that
characterizes real fluctuations. The main differences seem to be
in the high frequency tail of artificial data where effects due
to aliasing, absent in the left--hand--side panel, can be noticed.
\section{Results of the numerical simulations of parametric instability}

Recently, \cite{pri99, mal00, mal01, pri03} investigated in
detail how the parametric instability could be responsible for
typical features observed in the radial evolution of the
Alfv\'enic turbulence in the solar wind high speed streams. This
instability develops in a compressible plasma and, in its
simplest form, involves the decay of a large amplitude Alfv\'en
wave (generally called ``pump wave'', or ``mother wave'') in a
magnetosonic fluctuation and a backscattered Alfv\'en wave. The
wavevectors and frequencies of the fluctuations generated in this
process are mutually related through well precise ``resonance
conditions'' \citep{sag69}. This mechanism can be viewed as a way
for decorrelating an initially coherent state (the large
amplitude mother Alfv\'en wave). In fact, a circularly polarized
Alfv\'en wave is an exact solution of the ideal
magnetohydrodynamics equations even in the compressible case.
Hence, it should propagate unperturbed in a uniform plasma.
However, in presence of even very small perturbations in density,
this wave is subject to the parametric instability and it decays
producing fluctuations of different kinds.In fact, in slow
streams, the inhomogeneities of the background magnetic structure
supply a source of decorrelation for the Alfv\'en waves coming
from the sun (\cite{bru85}). On the converse, in fast streams,
where the magnetic field is more homogeneous, waves should travel
almost undisturbed and the observed (although slower than in slow
wind) radial evolution of the Alfv\'enic turbulence need some
mechanism to be ascribed to. Although, it has been shown
(\cite{rob92, gold95}) that plasma instabilities generated by
velocity shears play a relevant role in the radial evolution of
turbulence, another possibility is represented by parametric
instability as shown by \cite{mal00} and \cite{pri03}. These last
studies related to parametric instability focus the attention on
the effects of this instability on the evolution of a large
amplitude, circularly polarized, non-monochromatic Alfv\'en wave
in a one dimensional case. The spectrum of this initial
perturbation had a break at a certain wavelenght, like the
spectrum of the Alfv\'enic fluctuations coming from the sun. They
found that the Alfv\'enic correlation of the initial perturbation
is lost during the time evolution, because of the parametric
instability, leading to a production of both backscattered
Alfv\'enic perturbations and magnetosonic waves. Finally, these
perturbations evolve non linearly, producing approximately power
law spectra and a reduction of the normalized cross helicity. The
results found are qualitatively in good agreement with solar wind
observations carried out by several authors (see review by
\cite{tu95}). In addition, \cite{mal00} observed that the
turbulent development of the instability leads to the formation
of shock waves and to an intermittent behaviour of the
dissipation. In particular, looking at the evolution of the
flatness of velocity and magnetic field fluctuations,
\cite{pri03} found a good qualitative agreement of the results of
the simulations with the analysis of the same quantities
performed by \cite{bru03a}. It is then natural, in order to offer
a possible different interpretation of the results shown in the
previous sections, at least those concerning fast solar wind
streams, to see whether turbulence induced by parametric
instability has characteristics similar to those described in the
solar wind in the previous sections. To accomplish this aim, we
furtherly analysed the results of the numerical simulations
described in \cite{pri03}. The details of the numerical code can
be found in \cite{pri99}, \cite{mal00} and \cite{mal01}, whilst
further details concerning the simulations are given in
\cite{pri03}.

We simulate the evolution of a broad--band Alfv\'enic fluctuations
in a compressible plasma, during their outwards propagation in the
heliosphere. Similarly to the in--situ observations, the initial
spectrum has a break--point. During the run of the simulation,
inward propagating fluctuations start to appear and form a
power--law spectrum at small $k's$. As already pointed out by
\cite{tu89}, this feature might suggest that parametric decay
mechanism is at work in the solar wind.

The simulation domain is one-dimensional, periodic and we use
cartesian geometry.
The reference frame is chosen in such a way that the initial
Alfv\'en wave is circularly polarized in the $x-z$ plane and it
propagates along the $y$ direction. A background constant
magnetic field intensity ${\bf B}_0$ is imposed in the
propagation direction of the wave: the resulting total field has,
therefore, uniform intensity everywhere.
The homogeneous boundary conditions limit the application of this
study to the fast wind, where the background magnetic field is
rather homogeneous. In our framework, the time evolution of the
quantities represents the radial evolution of the fluctuations in
the solar wind, while the spatial variations are the numerical
counterpart of samples of the observed data at a given distance
from the sun. We study the evolution of the parametric
instability for 180 $\tau_A$ ($\tau_A$ is the Alfv\'en time based
on the initial background radial magnetic field and density, i.e.
the time needed for the wave, whose wavelength is the largest in
our spectral domain, to go across the simulation box). In the rest
of the paper, we plot quantities at time $t_1 = 45 \tau_A$ and
$t_2 = 180 \tau_A$, the former corresponding to a time much
before the saturation of the instability, the latter to a time
longer than the saturation time, which is reached at $t_{sat}\sim
100 \tau_A$. Practically, we consider a situation in which the
instability has only weakly taken place and another in which it
has already completely developed, that should be representative
of the state of the solar wind closer to the sun and further away
from it, respectively.

We estimated the time needed by the instability to saturate and to
reproduce the spectral features observed at 0.9 AU. We found that
a period of time between 6 and 7 days is necessary. This
estimation, although longer, is still within the same order of
magnitude of the expansion time required by the solar wind to
travel between 0.3 and 1 AU. The above evaluation is based on the
fact that between $t_1$ and $t_{sat}$ there are about $55\tau_A$.
In order to estimate $\tau_A$ we compared the frequencies
corresponding to the observed spectral break at 0.3 AU in Helios
data with the corresponding one shown in our simulation at
$t=t_1$.

\begin{figure}[t]
\vspace*{2mm}
  \caption{\label{fig:fig017} The top panel of the figure refers to
  the results of the numerical simulations of parametric instability
  at time $t = t_1 = 45 \tau_A$, while the bottom panel refers to the
  the results at time $t = t_2 = 180 \tau_A$. Each point of both plots
  represents the location of the tip of the magnetic field vector in a
  point of the simulation domain, in the same format of
  Fig.~\ref{fig:fig002}, which the present figure should be compared with.
  The difference in the shape of the plots, cylindrical for
  this figure, spherical for Fig.~\ref{fig:fig002} is explained in the
  text. Also in this case, the simulated data at $t = t_1$, corresponding
  to the real data at $0.3 AU$, have a more uniform distribution
  on the cylindrical surface, whilst the data at $t=t_2$ (to compare with
  real data at $0.9 AU$) show long jumps, followed by small
  amplitude oscillations around the new values.}
\end{figure}
In Fig. \ref{fig:fig017}, we show the three-dimensional plots of
the magnetic field components, normalized to the average magnetic
field intensity, at the time $t_1$ (upper panel) and $t_2$ (lower
panel), respectively. These plots are to be compared to Fig.
\ref{fig:fig002}. One can see that at $t=45 \tau_A$ the tip of the
magnetic field vector (represented by the line) moves in space in
a rather uniform way and covers approximately the surface of a
cylinder. This behaviour is qualitatively very similar to that
shown in the upper panel of Fig. \ref{fig:fig002}, only, at
difference with that case, here the tip of the vector covers
almost uniformly a cylinder instead of a sphere. This difference
is due to the fact that the numerical simulation is
one-dimensional: this implies (due to the divergenceless
condition for the magnetic field) that the variations of the
magnetic field components are only orthogonal to the propagation
direction of the waves, whilst the parallel component remains
constant during the time evolution. Moreover, since the initial
perturbation is circularly polarized, the trajectory of the tip
of the vector would describe a circular line. In order to improve
the visualization of the curve, to make it three-dimensional
instead of two-dimensional, we replaced the constant component of
the magnetic field with a linearly growing function between zero
and one, that makes the tip of the vector stay on a cylindrical
surface. The real data are three-dimensional, instead, and the
approximately constant field intensity produces the spherical
pattern plotted in Fig.~\ref{fig:fig002}, as already explained in
section \ref{data}.

At the later time $t_2=180 \tau_A$ the situation changes
dramatically. The tip of the magnetic field vector describes a
more patchy pattern, characterized by large jumps, followed by
smaller fluctuations around a single direction, and so on. This
pattern is qualitatively similar to that observed in Fig.
\ref{fig:fig002}. In conclusion, results of parametric instability
simulation seem to account for the observed transition to a sort
of L\'evy walk observed in the real solar wind data (see also
\cite{bru04}).
\begin{figure}[t]
\vspace*{2mm}
  \caption{\label{fig:fig018} Results of the numerical simulations
  of the parametric instability at time $t = t_1 = 45 \tau_A$.
  Normalized vector displacements versus the spatial coordinate $y$
  are shown in the top panel while, normalized
  vector intensities are shown in the bottom panel.
  This figure should be compared with Fig.~\ref{fig:fig006}.
  The top panel shows large fluctuations which are basically uncorrelated
  with the profile of the magnetic field intensity in the bottom panel.
  This behaviour is qualitatively similar to that observed in the
  fast wind at $0.3 AU$ (see Fig.~\ref{fig:fig006})}
\end{figure}
\begin{figure}[t]
\vspace*{2mm}
  \caption{\label{fig:fig019} Normalized vector displacements and normalized
  vector intensity versus position in the simulation box are shown in the
  same format as of Fig.~\ref{fig:fig018} for the results of the numerical
  simulations at time $t = t_2 = 180 \tau_A$. At difference with the
  previous figure, vector displacements (top panel) appear to be more
  structured in space and some sort of correlation with the field
  intensity (bottom panel) can be recognized. This behaviour is
  qualitatively similar to that observed in fast solar wind at $0.9 AU$
  (Fig.~\ref{fig:fig007}).}
\end{figure}

As a second point of agreement with the observations, we plot the
vector displacements $|\delta\underline{B}|$, with respect to a
fixed direction, normalized to the mean magnetic field intensity
as a function of the independent variable $y$. We compute this
quantity by considering the magnetic field at a given time $t$
and evaluating the vector differences of its components with the
fixed direction $(0;1;0)$ in each simulation grid point. Since in
our simulation the background magnetic field has components ${\bf
B}_0 = (0;1;0)$, we are practically plotting the vector
displacements of the magnetic field fluctuations. Finally, we
compute the intensity of this vector in each point. The results of
this computation are shown in the upper panel of Fig.
\ref{fig:fig018} at the time $t_1=45 \tau_A$ and Fig.
\ref{fig:fig019} at the time $t=180 \tau_A$. Along with this
curves, the magnetic field intensities $|B(y)|$, normalized to
its mean value, at the same times, are shown in the lower panels
of the figures.  This graphics should be compared with the ones
in Fig. \ref{fig:fig006} and \ref{fig:fig007}, where the analogous
quantities are plotted for the fast solar wind at $0.3$ and $0.9
AU$, respectively.

Also in this case the qualitative similarity between the results
of the simulations and the observed solar wind data is
remarkable. At the time $t=t_1$, corresponding to $0.3 AU$, the
vector displacements have a quite random behaviour, although not
as ``noisy'' as in Fig. \ref{fig:fig006}, and no evident
correlation between the vector displacements and the magnetic
field intensity is observed. Note that the fluctuations of the
magnetic field intensity are rather small at this time, due to
the fact that the initial wave is circularly polarized.

On the converse, in Fig. \ref{fig:fig019}, the vector
displacements of the magnetic field appear to be more structured,
characterized by fast rotations of the vector, followed by
smaller oscillations around the new position. Moreover, there
exists a clear correlation between the strongest gradients of the
vector displacements and the ones of the magnetic field intensity,
as already observed in the real solar wind fast streams.
\begin{figure}[t]
\vspace*{2mm}
  \caption{\label{fig:fig020} PDFs of vector displacements
  $|\delta\underline{B}_\lambda|$, at the scale $\lambda = \lambda_{max}/628$,
  normalized to $\sigma$ for the results of the
  numerical simulations of the parametric instability at time
  $t = t_1 = 45 \tau_A$ and $t = t_2 = 180 \tau_A$ are
  shown in the top and bottom panels, respectively. The two thin solid curves
  refer to as many lognormals contributing to form the thick solid curve which
  best fits the distribution. Parameters relative to the fit are reported in
  Tab.~\ref{leo:tab}.}
\end{figure}

Finally, we plotted in Fig.~\ref{fig:fig020} the PDFs of vector
displacements $|\delta\underline{B}_\lambda|$, defined as
\begin{equation}\label{eq:leodeltab}\nonumber
|\delta\underline{B}_{\lambda}(x,t)|=\sqrt{\sum_{i=x,y,z}(B_i(x+\lambda,t)-B_i(x,t))^2}
\end{equation}
normalized to their standard deviations, at times $t_1$ and
$t_2$. In eq.\ref{eq:leodeltab}, $t$ represents a specific phase
of the evolution, $x$ a generic position within the simulation
box and $\lambda$ is the scale length analogous to the sampling
time of real data used to compute the PDFs in section \ref{data}.
However, at difference with real data, where the resolution of
$6~sec$ still lies in the inertial range, in the simulations we
have to keep into account the fact that the smaller scales are
affected by viscosity and diffusivity. Thus, we evaluated the
vector differences at a length scale $\lambda=\lambda_{max}/628$,
where $\lambda_{max}$ is the maximum wavelength excited at the
beginning of the simulation, and equals the length of the
simulation domain. This length scale is rather small, compared to
the integral scale of the domain, but still in the inertial range
of the spectrum.

The similarity with Fig.~\ref{fig:fig008}, where the PDFs of the
vector displacements $|\delta\underline{B}|$ are shown for the
fast wind at $0.3$ and $0.9 AU$, is evident. Also in this case,
it is possible to fit effectively the curves with two lognormal
distributions and the trend is similar to that observed in the
solar wind data. In fact, the two lognormal distributions have
comparable heights at $t=t_1$, while the population relative to
the long vector displacements increases its importance at
subsequent times ($t=t_2$). The parameters of the fits are shown
in table \ref{leo:tab}.

Another point of similarity with the analysis of real data
regards the power spectrum. As a matter of fact, as already shown
by \cite{mal00} and \cite{mal01}, the power spectrum obtained
from the trace of the spectral matrix of the Alfv\'enic
fluctuations, after the saturation of the parametric instability
has been reached, shows a clear evidence of a power law inertial
range similar to the one observed in Helios data at 0.9 AU.

%
\begin{table*}[t]
  \caption[]
{\label{leo:tab}Numerical simulation: Values of the Parameters
Obtained from the Fit of the PDF relative to vector
displacements} \vskip4mm
\renewcommand{\arraystretch}{1.2}
\begin{tabular}{cccccccc}
\hline\noalign{\vskip1mm} Simulated time & $r^2$ & $A_1$ &
$\sigma_1$ & $\delta_1$
& $A_2$ & $\sigma_2$ & $\delta_2$  \\
\hline $45 \tau_A $ & $0.954$ & $1.13\pm0.14$  & $1.43\pm0.09$ &
$1.08\pm0.20$
&  $0.47\pm0.06$  &  $0.35\pm0.03$  &   $0.94\pm0.03$  \\
$180 \tau_A $ & $0.996$ & $0.52\pm0.56$  &  $0.72\pm0.21$  &
$0.45\pm0.27$
&  $0.09\pm0.55$  &  $0.69\pm0.95$  &   $0.15\pm0.45$  \\
\noalign{\vskip1mm}\hline
\end{tabular}
\end{table*}
%

%
We like to stress that a direct quantitative comparison between
the real data and the simulated data is meaningless. In fact, a
direct comparison would require to fix appropriate length and
time scales in the simulation. Moreover, as described in
\cite{pri03}, the model used is very simple, due to the
one-dimensionality constraint, to the cartesian geometry and
especially to the fact that the solar wind is actually an
expanding medium. However, the results of the simulations show a
qualitative behaviour very similar to that observed in real data
of the fast solar wind. We conclude that the parametric
instability offers a possible alternative explanation of the
observed data.
\section{Conclusions}\label{sec:end}

In this paper we focused on the statistics followed by
interplanetary magnetic field fluctuations on a 6-sec time scale,
well inside the MHD regime (\cite{bav82}, as observed in solar
wind turbulence between 0.3 and $0.9$ AU. In particular, we aimed
to understand the spatio--temporal evolution of the magnetic
field vector through the study of changes experienced by both
vector orientation and intensity. Several previous works, which
dealt with a statistical approach to this same problem,
considered different aspects connected to directional
fluctuations as, for example, power associated to the
fluctuations, their radial evolution, their anisotropy, the
nature of the fluctuations, their generation mechanisms, and so
on but, none of them, to our knowledge, has ever studied how and
why the orientation of these fluctuations changes with time.
There have been only few attempts to study similar problems but
always limited to single case studies (\cite{nak89, tu91, tsu94,
ril96, bru01}. The most recent statistical approach to the same
problem is represented by a paper by \cite{bru04} in which these
authors concluded that the temporal evolution of magnetic field
and wind velocity vectors directions might follow a sort of
L\'evy walk. That paper, although based on larger time scales and
on a weak statistics, represents the first attempt to understand
the influence due to propagating modes and convected structures
on the orientation of velocity and magnetic field vectors within
MHD turbulence. Following this analysis and, using a more robust
statistics, we found that PDFs of interplanetary magnetic field
vector differences within high velocity streams can be reasonably
fitted by a double log-normal distribution. In other words,
vector differences, which are due to the two distinct
contributions of directional uncompressive fluctuations and
purely compressive fluctuations, can be separated in two distinct
PDFs. Moreover, the lognormal nature of the PDFs might suggest a
\textsl{multiplicative process} at the origin of these
fluctuations, that is typical of a turbulent cascade.
Furthermore, it only applies to definite positive quantities,
like the vector or angular displacements we analyze in this paper.
Incidentally, the \textsl{multiplicative cascade} notion was
introduced by Kolmogorov into his statistical theory
(\cite{kol41, kol62} of turbulence as a phenomenological
framework to accomodate extreme behaviour observed in real
turbulent fluids.

Another interesting feature of these distributions is that the
two PDFs have a different weight since one of them, the one that
represents the smallest $|\delta\underline{B}_i|$ is always
considerably smaller than the other one. Moreover, while the
smaller PDF does evolve with heliocentric distance, decreasing
its own relevance, the largest PDF seems to remain almost
unaffected. Now, if we consider, as already suggested
(\cite{bru03a}), MHD turbulence mainly due to propagating,
uncompressive fluctuations of Alfv\'enic origin and to convected
compressive structures, it comes natural to identify these
different contributions to turbulence with the two PDFs we found.
In addition, we would expect a different radial evolution since
only propagating Alfv\'enic modes, interacting non--linearly,
undergo a considerable turbulent evolution as we already know
from literature (\cite{tu95}. We found that the relative
contribution to the total PDF of what we identify with
uncompressive fluctuations varies from $34\%$ at $0.3$ AU to
$18\%$ to $0.9$ AU, in terms of relative probability. In other
words what we identify with the Alfv\'enic contribution results
to be somewhat smaller than the contribution due to the convected
structures. Similar conclusions were reached by \cite{bie96} who
gave an estimate for the Alfv\'enic component around $15\%$ of
the total power associated to turbulence. Thus, we might associate
our convected structures to the 2--D turbulence identified in the
solar wind by \cite{bie96} and \cite{mat90} who model
interplanetary magnetic turbulence as made of slab and quasi--2D
turbulence only. However, the dominant 2--D magnetic turbulence
is characterized by the fact that its wave vector results to be
normal to the ambient magnetic field direction. As a consequence,
we would expect to see a radial evolution even stronger than the
one we observed for the slab component which has its wave vectors
parallel to the ambient field. As a matter of fact, the turbulent
cascade acts preferably on wave numbers perpendicular to the
ambient magnetic field direction, as suggested by the three--wave
resonant interaction (\cite{she83, bon85}. On the contrary, the
dominant component of the turbulence observed by Helios is the
least affected by the radial evolution and, probably, should not
be identified with the 2--D turbulence. Another possibility is
that the 2--D turbulence is mixed together with the slab
turbulence and represented by the smaller PDF which experiences
the stronger radial evolution. If this is the case, our analysis
suggests that interplanetary fluctuations are made of three
rather than two components: slab, 2--D and convected structures
which would support the three component model by \cite{mat99}.
This view is corroborated by the fact that the PDF of
$|\delta\underline{B}_i|$ within slow wind can be fitted by a
single lognormal whose parameters only slightly change with
heliocentric distance. As a matter of fact, this behavior has to
be expected if we consider that slow wind is poor in Alfv\'en
modes and its turbulence is already fully developed by the time
we observe it at $0.3$ AU (\cite{tu90}. Consequently, between.3
and .9 AU, fluctuations do not undergo the same turbulent
evolution observed in the fast wind and the constancy of the PDF
of $|\delta\underline{B}_i|$ should be expected. We like to
stress that interplanetary observations revealed that slow wind
MHD fluctuations are intrinsically different from those observed
in fast wind. In fact, \cite{bav00} showed that the Els\"asser
ratio $e^-/e^+$, which is always around $1$ within slow wind,
saturates to $\sim 0.5$ within fast wind at a distance of $\sim
2.5$ AU.

One more interesting observation regards the topology showed by
these fluctuations within fast and slow wind. We showed that the
trajectory followed by the tip of the magnetic vector during its
turbulent fluctuations follows a structured path. This path
appears more clearly when the PDF of $|\delta\underline{B}_i|$ can
be fitted by a single lognormal, as in the case of slow wind
regardless of heliocentric distance. However, within fast wind
this structured path can be more easily observed with increasing
the heliocentric distance, in concurrence with the depletion of
the Alfv\'enic fluctuations. In other words, Alfv\'enic modes
mask the underlying magnetic, quasi--static structure convected
by the wind. The superposition of these two types of fluctuations
is such that the final motion is characterized by extreme
behaviour. Referring to the 3D representation used in this paper,
the tip of the vector appears to be trapped within a certain
solid angle for sometime but, occasionally, it escapes this
limited angular region and quickly travels, in a few time steps,
to finally end up in another angular region characterized by a
different average orientation. These large jumps should be
accounted for by the larger PDF and should be related to similar
large jumps studied by \cite{bru01} and interpreted as tangential
discontinuities marking the border between adjacent flux tubes.
On the contrary, local fluctuations, clustering around certain
average directions, should have Alfv\'enic nature and should be
identified by the smaller PDF. These results support and further
corroborate the recently re--proposed \textsl{spaghetti--like}
structure model (\cite{bru01} firstly introduced, although in the
context of cosmic ray modulation, by \cite{mcc66} to describe
interplanetary magnetic field topology.

Finally, adopting a sort of feedback procedure, we cross--checked
the soundness of our fitting scheme showing that artificial data
obtained from the tip of a vector that randomly walk on the
surface of a sphere of constant radius, performing directional
jumps which obey to a double lognormal, provides results similar,
in some aspects, to those observed in interplanetary space.

However, the interplanetary observations we have do not allow to
understand whether these structures come directly from the Sun or
are locally generated by some mechanism. Recent theoretical
results by \cite{pri03} showed that coherent structures
responsible for the radial dependence of Intermittency as
observed in the solar wind (\cite{bru03a}, might be locally
created by parametric decay of Alfv\'en waves. These authors,
showed that during the turbulent evolution, coherent structures
like shocklets and/or current sheets were continuously created
when the instability was active.
In order to see whether a similar mechanism may account for the
observed behaviour of the vector displacements and their
statistics, we further analyzed in this paper the results of the
simulations performed in \cite{pri03}. The results of this
investigation show a fairly good agreement, at least under the
qualitative point of view, between the simulations and the solar
wind data: either the evolution of the tip of the magnetic field
vector, or the correlation between the vector displacement at a
given scale with the magnetic field intensity fluctuations, or the
evolution of the PDFs of the vector displacement in time, all
show trends similar to those observed in the real fast solar wind
data. Unfortunately, a direct quantitative comparison between the
simulations and the data is difficult due to the limitations of
the model.
However, this mechanism, which might be active within fast wind,
should be less effective within slow wind given the remarkable
decoupling between magnetic field and velocity field within this
type of wind (\cite{kle93}.
Nevertheless, the enticing nature of the parametric instability
in explaining the results comes from some well defined fact: {\em
a}) it is a well defined mechanism of physical origin that
induces a turbulent evolution in the plasma and not an undefined
turbulence model; {\em b}) it is likely applicable to explain
many general observed features of fast solar wind, like the
evolution of the spectra, the decrease of the Alfv\'enic
correlation during the propagation in the heliosphere, and so on
(\cite{pri03}; {\em c}) the observed evolution of the vector
displacements and of their relative PDFs can be seen as a natural
consequence of the formation of shocklets and discontinuities in
the wind, organized in a sort of coherent structures, that
explain the long jumps observed in the magnetic field and the
structures in the vector displacements at larger distances from
the sun. In particular, the decrease with distance of the
lognormal component of the PDFs correlated to the Alfv\'enic part
of the turbulence, can be seen as the continuous transfer of
energy between the Alfv\'enic and magnetosonic components of the
waves during the evolution of the instability. However, a
definitive conclusion about this point needs further
investigations.

Another recent theoretical effort by \cite{cha04} models MHD
turbulence in a way that tends to the view and interpretation of
the interplanetary observations we presented in this paper, that
is the existence of two different components both contributing to
turbulence. The theoretical model presented by these authors
tells us that propagating modes and coherent, convected
structures are both necessary, inseparable ingredients of MHD
turbulence since they share a common origin within the general
view described by the physics of complexity (\cite{cha99, vaho04,
vas04}. Propagating modes experience resonances which generate
coherent structures which, in turn, will migrate, interact and
eventually generate new modes.

These theoretical models, which favour the local generation of
coherent structures, fully complement the possible solar origin of
the convected component of interplanetary MHD turbulence.

\begin{acknowledgements}
      Magnetic field 6sec averages derive from the Rome--GSFC magnetic
experiment onboard Helios 2 s/c. (PIs of the experiment were F.
Mariani and N.F. Ness)
\end{acknowledgements}


\end{document}